\documentclass[12pt]{article}%
\usepackage{amsmath}
\usepackage{amsfonts}
\usepackage{amssymb}
\usepackage{graphicx}%
\begin{document}

\title{Magnetic flux noise in the three Josephson junctions superconducting ring}
\author{E. Il'ichev~${}^{1}$ and A.N. Omelyanchouk~${}^{2}$ \\
{\small {\em ${}^{1}$ Institute of Photonic Technology,}} \\
{\small {\em Albert-Einstein str. 9, 07745, Jena, Germany}} \\
{\small {\em ${}^{2}$ B.Verkin Institute for Low Temperature Physics and
Engineering,}} \\
{\small {\em National Academy of Sciences of Ukraine,}} \\
{\small {\em 47 Lenin Ave., 61103 Kharkov, Ukraine}}}
\maketitle
\date{}

\begin{abstract}
We analyze the influence of noise on magnetic properties of a
superconducting loop which contains three Josephson junctions.
This circuit is a classical analog of a persistent current (flux)
qubit. A loop supercurrent induced by external magnetic field at
the presence of thermal fluctuations is calculated. In order to
get connection with experiment we calculate the impedance of the
low-frequency tank circuit which is inductively coupled with a
loop of interest. We compare obtained results with the results in
quantum mode - when the three junction loop exhibits quantum
tunneling of the magnetic flux. We demonstrate that the tank-loop
impedance in the classical and quantum modes have different
temperature dependence and can be easily distinguished
experimentally.
			85.25.Hv Superconducting logic elements and memory devices; microelectronic circuits.

\end{abstract}
\newpage
\section{Introduction}
Magnetic flux quantization in superconductors is used,
in particular, for realization of very sensitive magnetometers.
One of them is so-called radio-frequency (rf) SQUID \cite{bar}. The sensor of
rf SQUIDs is a single junction interferometer - a Josephson
junction which is incorporated in a superconducting ring with a
sufficiently small inductance $L$. When an external flux $\Phi_e$
is applied to an interferometer loop, the circulating supercurrent $I$
is induced and a flux $\Phi_i$ is admitted into the ring:
\begin{equation}
\Phi_i =  \Phi_e - IL.
\label{Eq:int}
\end{equation}
The phase difference $\varphi$ across a Josephson junction equals to a
normalized magnetic flux in the interferometer loop:
\begin{equation}
\varphi = 2\pi\frac{\Phi_i}{\Phi_0}+2\pi n. \label{Eq:phi}
\end{equation}
where $\Phi_0$ is the flux quantum, and $n$ is an integer. Since
the  Josephson current is related to the phase difference $\varphi$:
\begin{equation}
I = I_{c} sin \varphi, \label{Eq:Jh}
\end{equation}
where $I_{c}$ is the critical current, Eq.~\ref{Eq:int} can be
rewritten:
\begin{equation}
\varphi =  \varphi_e - \beta sin \varphi \label{Eq:phi1}
\end{equation}
where $\varphi_e = 2\pi {\Phi_e}/{\Phi_0}$ is the normalized
external flux and the constant
\begin{equation}
\beta = 2\pi LI_c /\Phi_0 \label{Eq:beta}
\end{equation}
is the normalized inductance of the interferometer.

From Eq.~\ref{Eq:phi1} it is clearly seen that the magnetic
properties of a single junction interferometer are defined by
parameter $\beta$. If $\beta < 1$ the $\varphi_{e}(\varphi)$
dependence is unique (see Fig.~1) and corresponding mode of SQUID
operation is so-called nonhysteretic. If $\beta > 1$ the
$\varphi_{e}(\varphi)$ dependence is multivalued (see Fig.~1) and
corresponding mode of SQUID operation is hysteretic.

A rf SQUID basically consists of a sensor (usually a single
junction interferometer) inductively coupled to a
radio-frequency-biased tank circuit. The flux $\varphi_{e}$ applied
to the sensor changes the effective inductance (or/and the
effective resistance) of the tank-sensor arrangement. Thus, a flux
change can be detected as changes in phase (or/and amplitude) of
the voltage across the tank circuit.

The classical mode for the single junction interferometer as well
as for corresponding rf SQUID in the presence of fluctuations have been
investigated in detail theoretically as well as experimentally
[2-7]. On the other hand the quantum mode for this
device is difficult to realize. Since an interferometer should be
hysteretic, it requires the finite $LI_c$ product, and therefore,
the finite coupling with environment. In order to avoid this
problem, a substitution of the geometrical inductance to the
Josephson one has been proposed. Indeed, if the number of Josephson
junctions in the loop $m > 2$ and for suitable junctions
parameters, a double degenerated state exists at any geometrical
inductance $L$. One of the simplest realizations here is a three-
junction interferometer, which is called also a persistent current (or
flux) qubit \cite{Moj}. Such qubit was fabricated by several teams
and quantum regime was convincingly demonstrated.

If external magnetic flux is
\begin{equation}
\varphi = \pi+2\pi n. \label{Eq:ext}
\end{equation}
the hysteretic interferometer exhibits double-degenerated energy
states, see Fig.~3. These states correspond to the different
directions of the interferometer current. If temperature is low
enough and for suitable parameters of Josephson junctions the
magnetic flux can tunnel between the two potential minima. Below
we will call the systems under consideration quantum if in their
dynamics there is quantum tunneling. If, for some reasons, the
quantum tunneling is suppressed, we will call these systems
classical ones.

It is clear that the presence of the quantum tunneling ensures the
absence of the hysteresis at the $\varphi_{e}(\varphi)$
dependence. On the other hand jumps between two energy minima can
be originated by the fluctuations and the hysteresis will be washed
out. Therefore for both cases considered above the mode of the rf
SQUID operation will be nonhysteretic. It arises a natural
question: by analyzing a SQUID output signal is it possible to
distinguish between quantum mode (interferometer with "quantum
leak") and classical mode (interferometer in presence of the
fluctuations)? We address this paper to that question.

\section{Classical mode of a flux qubit in the presence of fluctuations}
The studied system presents the superconducting circuit (ring)
with three Josephson junctions, see Fig.~2. We consider the case
of small self inductance of the ring $L\rightarrow0$, therefore $
\Phi_e =\Phi_{i}$. The phases across each junction in the qubit
loop $\varphi_{i}$ satisfy to

\[
\varphi_{1}+\varphi_{2}+\varphi_{3}=\varphi_{e}.
\]

In the frame of RSJ model for Josephson junctions the current
through each junction is:

\begin{equation}
I=\frac{\hbar C_{i}}{2e}\frac{d^{2}}{dt^{2}}\varphi_{i}+\frac{\hbar}{2eR_{i}%
}\frac{d}{dt}\varphi_{i}+I_{ci}\sin\varphi_{i}+\delta
I_{i}(t),i=1,2,3 \label{Eq:RSJ}
\end{equation}
where $\hbar$ is the Plank constant, $e$ is the electron charge,
$C_i$ and $R_i$ are the junctions' capacitances and resistances
respectively. We restrict ourself to a practical case, when two
junctions in the loop are identical:
$C_{1}=C_{2}=C,I_{c1}=I_{c2}=I_{c},R_{1}=R_{2}=R$ and the third
junction has slightly smaller critical current (with the same
critical current density) $I_{c3}=\alpha I_{c}, 0.5 <\alpha <1$
and therefore $C_{3}=\alpha C, R_{3}=R/\alpha$. The presence of
the 'white noise' is given by the fluctuation currents $\delta I_{i}(t)$ with
correlator  $<\delta I_{i}(t)\delta I_{i}(t^{^{\prime}})>=2kT/R_{i}\delta(t-t^{^{\prime}})$ and mean value  $<\delta I_{i}(t)>=0$.

In dimensionless units:%

\begin{equation}
\omega_{R} = 2eRI_{c}/\hbar,\omega_{R}t=\tau\\
\end{equation}
and for negligible capacitance, Eq.~\ref{Eq:RSJ} can be rewritten:
\begin{equation}
I/I_{c} = d\varphi_{1,2}/d\tau+\sin\varphi_{1,2}+\delta\varphi_{1,2}%
=\alpha
d\varphi_{3}/d\tau+\alpha\sin\varphi_{3}+\delta\varphi_{3}(\tau).
\label{Eq:RSJ1}
\end{equation}
The correlators of $\delta\varphi_{i}$ are:
\begin{equation}
<\delta\varphi_{i}(\tau)\delta\varphi_{i}(\tau^
{^{\prime}})>=2D\delta_{i} (\tau
-\tau^{^{\prime}}),%
\end{equation}
where $D=kT/E_{J}$, $E_J=\frac{\hbar I_c}{2e}$.

By introducing the phases $\theta$ and $\chi$

\begin{align*}
2\theta = \varphi_{1}+\varphi_{2}, 2\chi =
\varphi_{1}-\varphi_{2},
\end{align*}
and taking into account that $\varphi _{3}=\varphi_{e}-2\theta$,
Eqs.~\ref{Eq:RSJ} can be presented in the form:

\begin{align*}
d\chi/d\tau &  =-\cos\theta\sin\chi+1/2(\delta\varphi_{2}-\delta\varphi_{1}),\\
(1+2\alpha)d\theta/d\tau   &
=-\sin\theta\cos\chi+\alpha\sin(\varphi_{e}
-2\theta)+\delta\varphi_{3}(\tau)-1/2(\delta\varphi_{2}+\delta\varphi_{1}).
\end{align*}

These equations can be reduced to:
\begin{align}
d\chi/d\tau &  =-\frac{\partial U}{\partial\chi}+\delta\chi(t),\\
d\theta/d\tau &  =-\frac{1}{(1+2\alpha)}\frac{\partial
U}{\partial\theta}+\delta\theta(t) \label{Eq:RSJ5}
\end{align}
where
\begin{equation}
U(\theta,\chi) = -\cos\theta\cos\chi-\frac{1}{2}\alpha\cos(\varphi
_{e}-2\theta) \label{Eq:pot}
\end{equation}
is the effective potential and the random forces are:
\[
<\delta\chi(\tau)\delta\chi(\tau^{^{\prime}})>=D\delta(\tau-\tau^{^{\prime}%
}),<\delta\theta(\tau)\delta\theta(\tau^{^{\prime}})>=\frac{3}{(1+2\alpha)^2}D\delta(\tau
-\tau^{^{\prime}}).
\]

The Langevin equations (11),(12) describe the random motion of the
'particle' with coordinates ($\theta$, $\chi$) in the periodic
potential (13), which is a set of bistable cells (eight-shaped
contours in Fig.~3a). We have numerically integrated these
stochastic equations by Ito's method (see \textsl{e.g.}
\cite{Gard}) for different values of parameter $\alpha$ and the
strength of the fluctuations $D$.  The typical traces of
$\theta(\tau)$ and $\chi(\tau)$ are shown in Fig.~4. They
correspond to random motion in bistable potential, Fig.~3. The
arrows indicate the switching from one unit cell in Fig.~3a to
another. With knowledge of $\theta(\tau)$ and $\chi(\tau)$ the
average circulating current in the ring is obtained as:
 \begin{equation}
I(\phi_{e})=I_{c}<<sin(\chi+\theta)>>.
\end{equation}

The averaging $<<...>>$ includes for each value of flux $\phi_{e}$
the average over time of traces ($\theta(\tau),\chi(\tau)$) and
the average over set of 50 traces.

The calculated in such way the current-flux curves for different
values of $D$ and different values of parameter $\alpha$ are
presented in Fig.~5 and Fig.~6.

From Eqs.~(11-13) the Fokker-Plank equation for distribution
function $P(\chi,\theta)$ can be reconstructed:
\begin{equation}
\frac{\partial P}{\partial
\tau}=\frac{\partial}{\partial\chi}(\frac{\partial U}{\partial
\chi}P)+\frac{D}{2}\frac{\partial^{2}}{\partial\chi^{2}}P +
\frac{1}{1+2 \alpha}\frac{\partial}{\partial\theta}(\frac{\partial
U}{\partial \theta}P)+\frac{1}{2}\frac{3
D}{(1+2\alpha)^{2}}\frac{\partial^{2}}{\partial\theta^{2}}P
\label{Eq:fok}
\end{equation}

The Fokker-Plank equation (15) admits the stationary potential
solution (see \cite{Gard}) in the special case $\alpha=1$,
\textit{i.e.} when all three junctions are identical. For
$\alpha=1$ the analytical solution reads:
\begin{equation}
P=\frac{e^{-\frac{2}{D}U(\chi,\theta)}}{\textsl{N}},
\end{equation}
\begin{equation}
\textsl{N}=\int\int d\chi d \theta e^{-\frac{2}{D}U(\chi,\theta)}
\end{equation}

Since the potential $U$ is $2\pi$ periodical function of variables
$\chi$ and $\theta$ the average current in the ring is:
\begin{equation}
\frac{I}{I_c}=\frac{\int^{2 \pi}_{0}\int^{2 \pi}_{0}d\chi
d\theta sin(\chi+\theta)e^{-\frac{2}{D}U(\chi,\theta)}}{\int^{2
\pi}_{0}\int^{2 \pi}_{0 }d\chi d\theta
e^{-\frac{2}{D}U(\chi,\theta)}} \label{Eq:Ifok}
\end{equation}

In Fig.~7 we compare the numerical results (circles) and the
$I(\phi_{e})$ obtained from the analitical formula (18)  (solid
line) for the case $\alpha=1$ and $D=0.2$. This comparison was
used as an additional calibration of our numerical procedure,
which is working at arbitrary values of $\alpha$.

\section{The probing of qubit's state in classical and quantum modes.}

    As we wrote in introduction we probe a qubit with making use of a
tank circuit by the impedance measurement technique \cite{fnt}. It
was convincingly shown \cite{fnt,Shev} that the observable phase
difference $\delta(\varphi_{e})$ between tank current $I_{rf}$ and
tank voltage $V_{rf}$ both in classical and quantum modes reads:

 \begin{equation}
 \tan\delta(\varphi_{e})\approx \Theta \frac{dI}{d\varphi_{e}},
\end{equation}
where $\Theta$ is \textsl{const}, which characterizes the
inductive coupling of the qubit with the tank circuit. By using the
results of Section 2 we calculated output signal
$\delta(\varphi_{e})$ for classical noise affected  mode. For
different levels of noise $D$ (all of those correspond to
nonhysteretic regime) the dependencies $\delta(\varphi_{e})$ are
shown in Fig.~8.

In the quantum noise free mode (see Appendix A) the nonhysteretic
behavior is achieved by the tunneling between two wells. The phase
shift $\delta(\varphi_{e})$ in this case is described by Eqs.
(19,22). It is presented in Fig.~8 for the same as in the classical
case value of the qubit-tank coupling constant $\Theta$ and
experimentally realized qubit's parameters
$I_{p}\Phi_{0}=200 GHz, \Delta=1.5 GHz$.

Comparing classical and quantum modes (Figs.~8,9) we have found that
in quantum mode the dip on the $\delta(\varphi_{e})$ dependence
remains constant at wide temperature range $kT \leq \Delta$. This
reflects the fact that tunneling splitting $\Delta$ does not
depend on the temperature. The depth of the dip is changed with
temperature - excitations to the upper level depress the value of
the average qubit's current. For classical mode the situation is
rather different. First of all for reasonable set of qubit
parameters it is impossible to get such narrow and profound dip
similar to obtained in quantum mode. Moreover, the temperature
dependence of $\delta(\varphi_{e})$ dip demonstrates that its
width strongly depends on $T$. Therefore by analyzing the
temperature dependence of $\delta(\varphi_{e})$ one can easily
distinguish between quantum and classical modes.

In conclusion we analyzed the temperature dependence of imaginary
part of impedance for three-junction loop-tank circuit arrangement
in quantum and classical modes. We argued that impedance for these
modes have quite different temperature dependencies and, therefore,
can be easily distinguished experimentally.

\appendix

\section{ Quantum mode of a flux qubit}
Since the tunnel splitting in flux qubit is much smaller than the
difference between upper energy levels, qubits are effectively
two-level quantum systems. In the two level approximation a flux qubit
can be described by the pseudo-spin Hamiltonian
\begin{equation}
H(t)=-\Delta \sigma _{\!x}-\varepsilon \sigma _{\!z}\;,
\label{eq01}
\end{equation}%
where $\sigma _{x},\sigma _{z}$ are the Pauli matrices; $\Delta $
is the tunneling amplitude. The qubit bias is given by
$\varepsilon = I_{p}\Phi _{0}f_{e}$, where $I_{p}$ is the
magnitude of the qubit persistent current and $f_{e}=\Phi
_{e}/\Phi _{0}-1/2$. The stationary energy levels can be easily
found from the Hamiltonian (\ref{eq01}):
\begin{equation}
E_{\pm }(\varepsilon)=\pm \sqrt{\varepsilon ^{2}+\Delta ^{2}},
\label{eq02}
\end{equation}
and the average value of the qubit current at temperature $T$ is:
\begin{equation}
I(\varepsilon)=\frac{\varepsilon
I_p}{\sqrt{\varepsilon ^{2}+\Delta ^{2}}} \tanh
\frac{\sqrt{\varepsilon ^{2}+\Delta ^{2}}}{kT},
\end{equation}

Note, that the dependence (21) is valid within the narrow interval
of $\Phi_{e}$ near $\Phi_{e}=\Phi_{0}/2$, where the potential $U$
(13) is bistable.

\newpage
FIGURE CAPTIONS \\
 Figure 1. $\phi(\phi_{e})$ for rf SQUID in nonhysteretic and hysteretic modes \\
 Figure 2. The scheme of persistent current (flux) qubit \\
 Figure 3. a. Contour plot of the potential $U(\theta,\chi)$ (13),
$\alpha=0.8$.  b. Bistable potential profile along line 1-2 in
Fig3a at $\phi_{e}=\pi$,$\alpha=0.8$. \\
 Figure 4. The random motion of phases $\theta$ and $\chi$ in
bistable potential (Fig.3b) for $\phi_{e}=\pi$,$\alpha=0.8$ and
$D=0.1$ \\
 Figure 5. The dependences $I(\phi_{e})$ for $\alpha=0.8$ and $D=0.5
(1),D=0.25 (2), D=0.1 (3)$ \\ 
 Figure 6. The dependences $I(\phi_{e})$ for $D=0.1$ and $\alpha=0.5
(1),\alpha=0.8 (2), \alpha=1 (3)$ \\
 Figure 7. The comparision of numerical (circles) and analitical
(solid line) calculations. $D=0.2$, $\alpha=1$ \\
 Figure 8. The phase shift $\delta(\varphi_{e})$ in classical mode.
$\alpha=0.8$ and $D=0.3 (1),D=0.2 (2), D=0.1 (3)$ \\
 Figure 9. The phase shift $\delta(\varphi_{e})$ in quantum mode.
$T/\Delta=0.1 (1),T/\Delta=1 (2)$ \\
\newpage
\begin{figure}
            \includegraphics[width=1\textwidth]{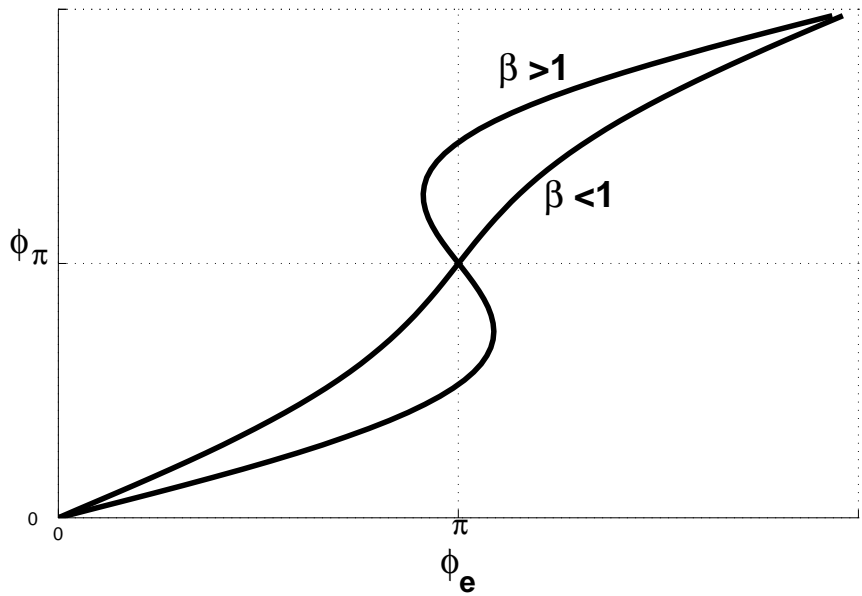}
        \caption{$\phi(\phi_{e})$ for rf SQUID in nonhysteretic and hysteretic modes}
\end{figure}

\newpage
\begin{figure}
            \includegraphics[width=0.8\textwidth]{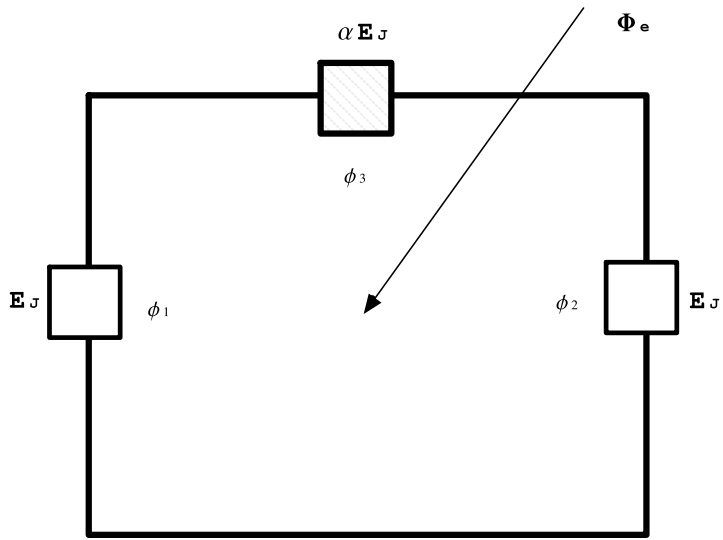}
        \caption{The scheme of persistent current (flux) qubit}
\end{figure}

\newpage
\begin{figure}

\includegraphics[width=1\textwidth]{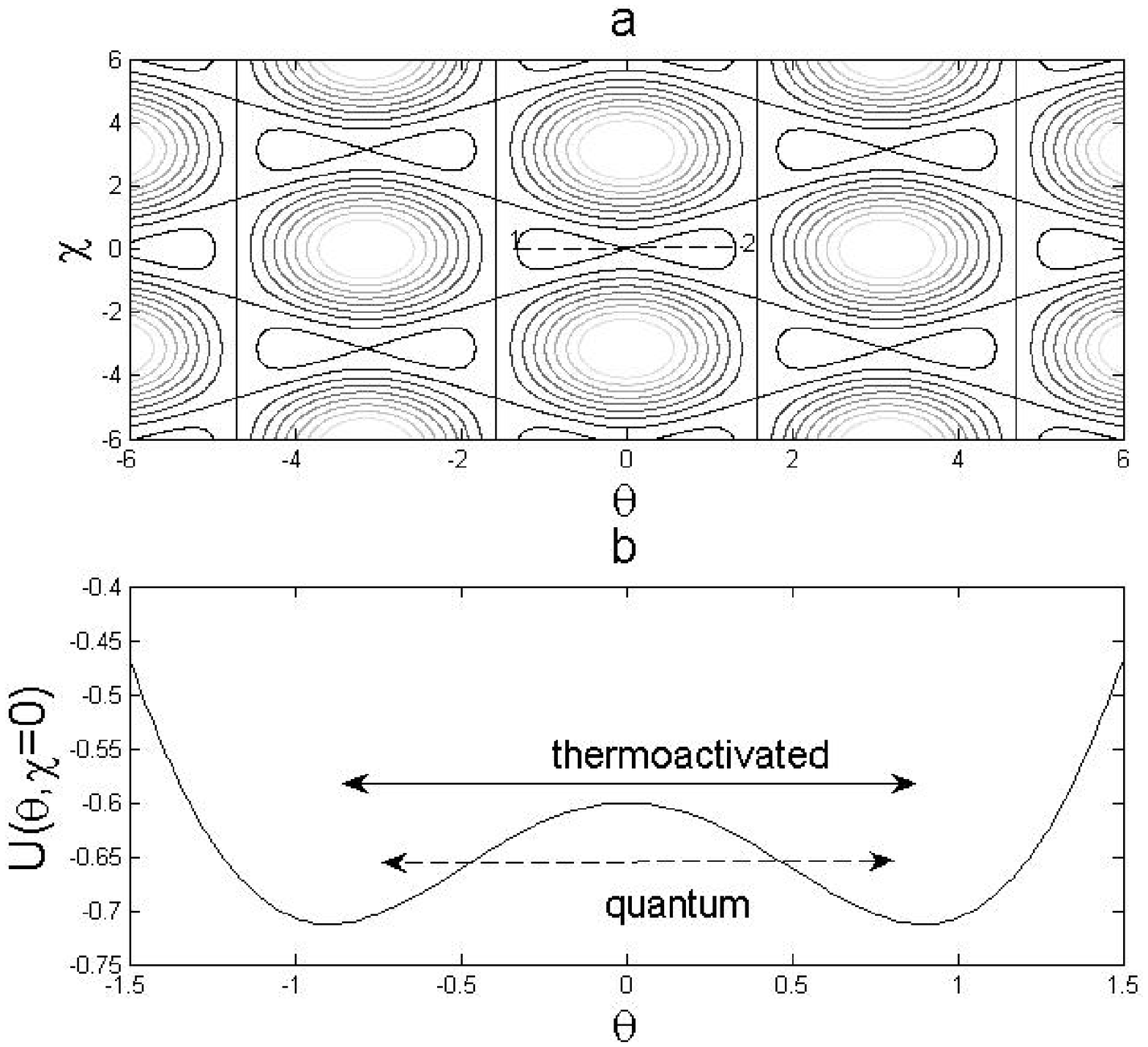}
\caption{a. Contour plot of the potential $U(\theta,\chi)$ (13),
$\alpha=0.8$.  b. Bistable potential profile along line 1-2 in
Fig3a at $\phi_{e}=\pi$,$\alpha=0.8$. }
\end{figure}

\newpage
\begin{figure}
            \includegraphics[width=1\textwidth]{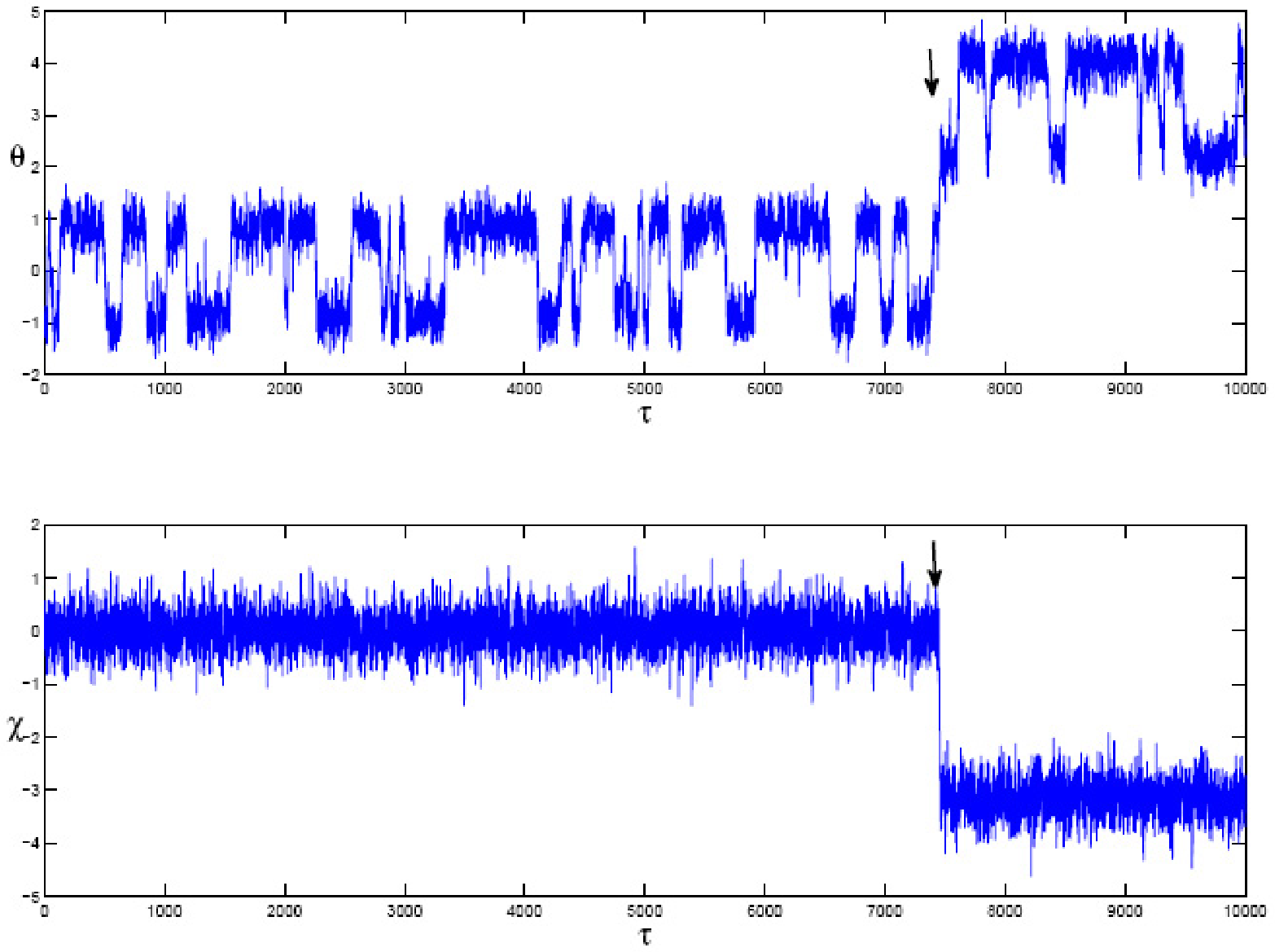}
\caption{The random motion of phases $\theta$ and $\chi$ in
bistable potential (Fig.3b) for $\phi_{e}=\pi$,$\alpha=0.8$ and
$D=0.1$}
\end{figure}
\newpage
\begin{figure}
            \includegraphics[width=1\textwidth]{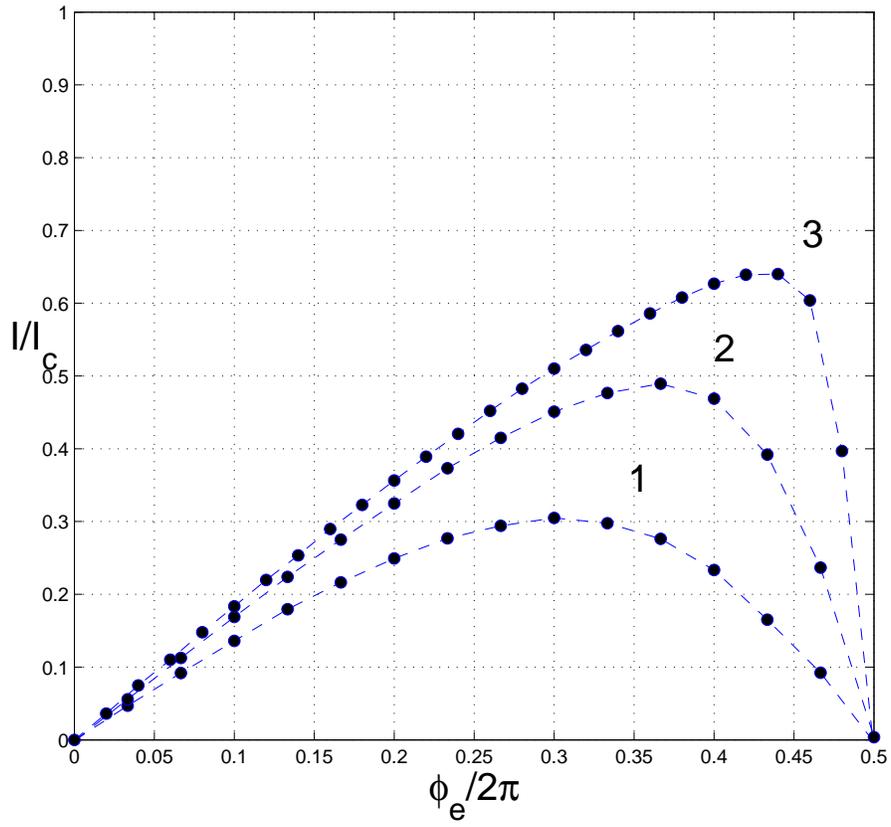}
\caption{The dependences $I(\phi_{e})$ for $\alpha=0.8$ and $D=0.5
(1),D=0.25 (2), D=0.1 (3)$ }
\end{figure}
\begin{figure}
            \includegraphics[width=1\textwidth]{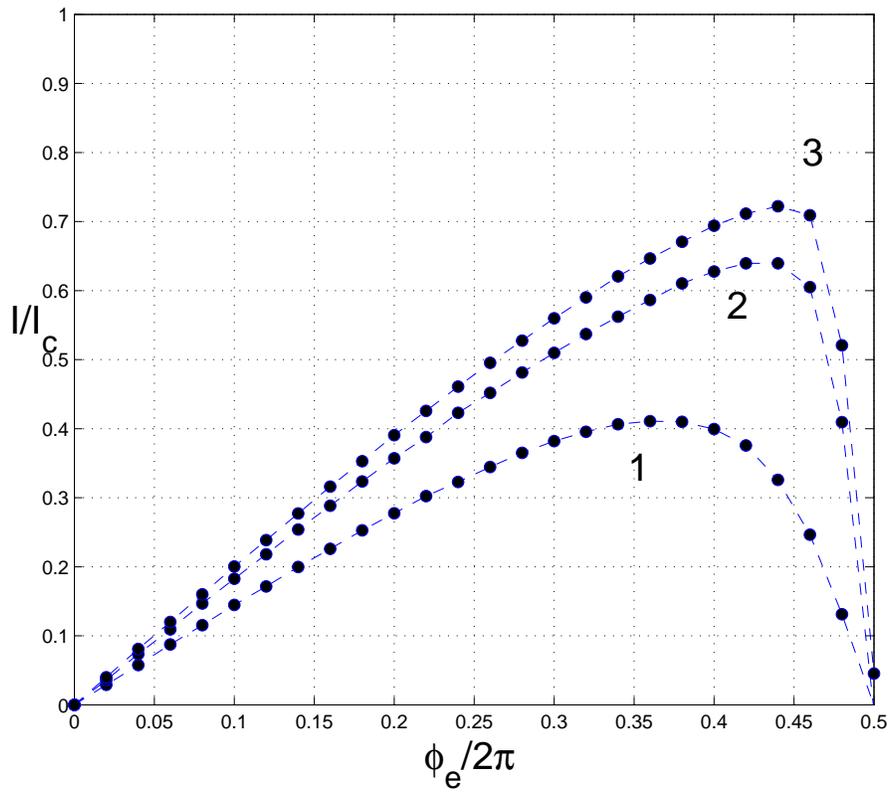}
\caption{The dependences $I(\phi_{e})$ for $D=0.1$ and $\alpha=0.5
(1),\alpha=0.8 (2), \alpha=1 (3)$ }
\end{figure}
\begin{figure}
            \includegraphics[width=1\textwidth]{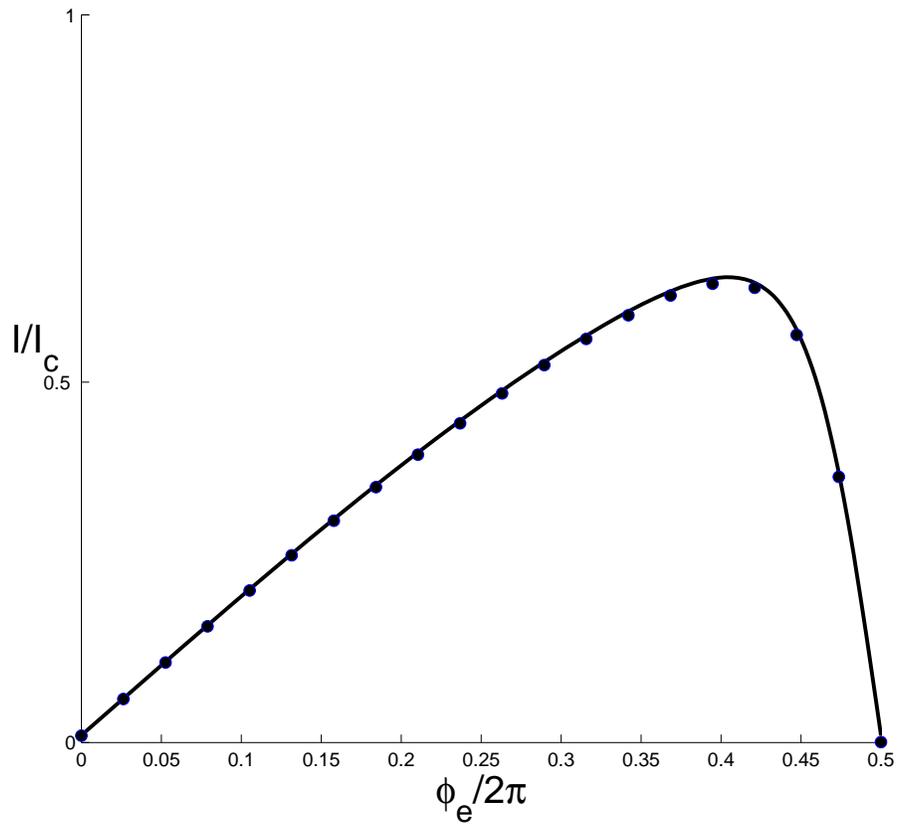}
\caption{The comparision of numerical (circles) and analitical
(solid line) calculations. $D=0.2$, $\alpha=1$ }
\end{figure}
\begin{figure}
            \includegraphics[width=1\textwidth]{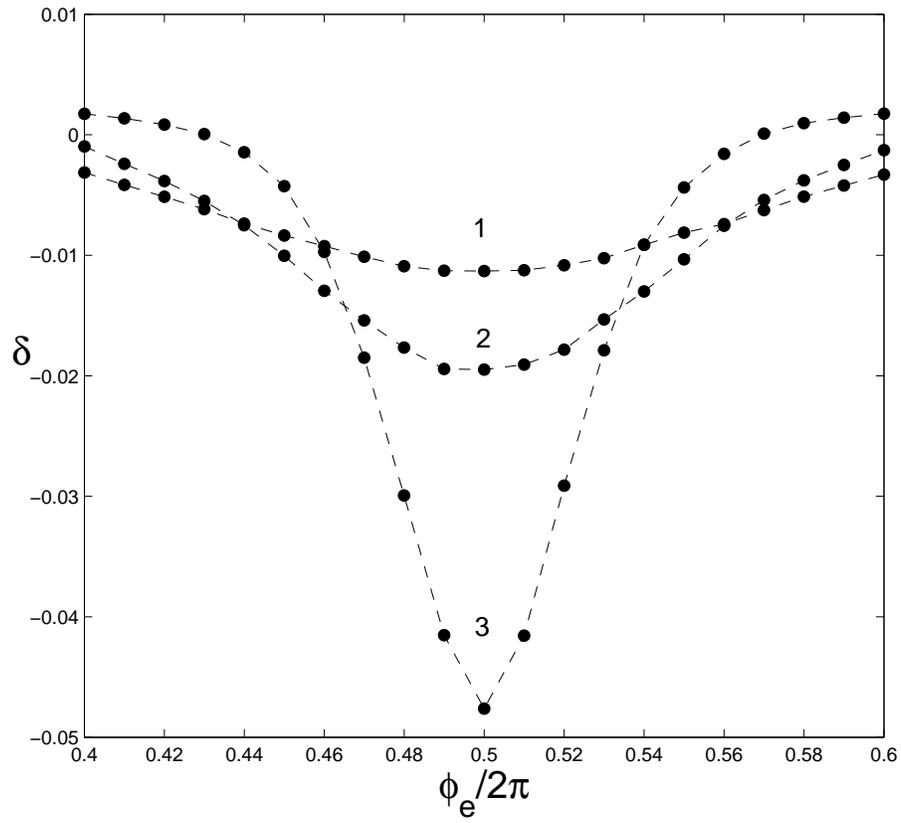}
\caption{The phase shift $\delta(\varphi_{e})$ in classical mode.
$\alpha=0.8$ and $D=0.3 (1),D=0.2 (2), D=0.1 (3)$ }
\end{figure}
\begin{figure}
            \includegraphics[width=1\textwidth]{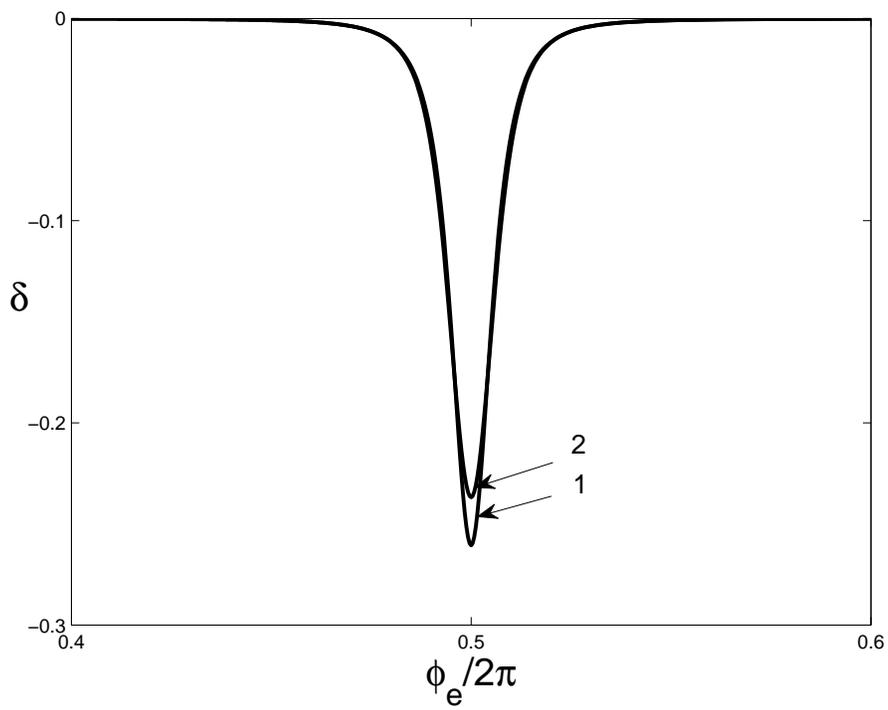}
\caption{The phase shift $\delta(\varphi_{e})$ in quantum mode.
$T/\Delta=0.1 (1),T/\Delta=1 (2)$}
\end{figure}
\end{document}